\documentclass[twocolumn]{aastex631}
\usepackage{amsmath}

\newcommand{\unit}[1]{\ensuremath{\, \mathrm{#1}}}

\graphicspath{{./}{Figures/}}

\received{XXX}
\revised{XXX}
\accepted{XXX}

\shorttitle{Planet Migration in 3D Radiative Disks}
\shortauthors{Yun et al.}

\begin{document}
\title{Effects of Radiative Diffusion on the Dynamical Corotation Torque in Three-Dimensional Protoplanetary Disks}

\author[0000-0003-4353-294X]{Han-Gyeol Yun}
\affiliation{Department of Physics \& Astronomy, Seoul National University, Seoul 08826, Korea}
\affiliation{SNU Astronomy Research Center, Seoul National University, 1 Gwanak-ro, Gwanak-gu, Seoul 08826, Republic of Korea}

\author[0000-0003-4625-229X]{Woong-Tae Kim}
\affiliation{Department of Physics \& Astronomy, Seoul National University, Seoul 08826, Korea}
\affiliation{SNU Astronomy Research Center, Seoul National University, 1 Gwanak-ro, Gwanak-gu, Seoul 08826, Republic of Korea}

\author[0000-0001-7258-770X]{Jaehan Bae}
\affiliation{Department of Astronomy, University of Florida, Gainesville, FL 32611, USA}

\author[0000-0002-2641-9964]{Cheongho Han}
\affiliation{Department of Physics, Chungbuk National University, Cheongju 28644, Republic of Korea}

\email{hangyeol@snu.ac.kr, wkim@astro.snu.ac.kr, jbae@ufl.edu, cheongho@astroph.chungbuk.ac.kr}

\begin{abstract}
The dynamical corotation torque arising from the deformation of the horseshoe orbits, along with the vortensity gradient in the background disk, is important for determining orbital migration rate and direction of low-mass planets. Previous two-dimensional studies predicted that the dynamical corotation torque is positive, decelerating the inward planet migration. In contrast, recent three-dimensional studies have shown that buoyancy resonance makes the dynamical corotation torque negative, accelerating the inward migration. In this paper, we study the dependence of the dynamical corotation torque on the thermal transport using three-dimensional simulations. We first show that our results are consistent with previous three-dimensional studies when the disk is fully adiabatic. In more realistic radiative disks, however, radiative diffusion suppresses the buoyancy resonance significantly, especially at high-altitude regions, and yields a positive dynamical corotation torque. This alleviates the issue of a rapid migration caused by the negative dynamical corotation torque in the adiabatic disks. 
Our results suggest that radiative diffusion together with stellar irradiation and accretion heating is needed to accurately describe the migration of low-mass planets.
\end{abstract}


\section{Introduction} \label{sec:intro}

Gravitational interactions between a protoplanetary disk and its embedded planet have important consequences in the evolution of the disk-planet system \citep{kn2012}. The planet induces gravitational wakes in the disk, and the gravitational interactions between the wakes and the planet promote angular momentum exchanges between them. If the total gravitational torque exerted on the planet by the wakes is negative (positive), the planet loses (gains) its orbital angular momentum and migrates radially inward (outward). The theory of the planet migration has brought many insights to the understanding of planet formation \citep{pj2018} and the architecture of observed planetary systems (e.g., \citealt{ba2014, ba2016, em2021}).

The planet migration is divided into three types depending on the planet mass. Low-mass (typically Earth-sized) planets experience type I migration which is mostly driven by spiral waves formed at the Lindblad resonances \citep{gt1980}.
The density wakes inside (outside) the orbital radius of the planet exert the positive (negative) torque, which is called the Lindblad torque. Previous studies have shown that the total Lindblad torque is negative for typical disk parameters, which results in an inward migration \citep{w1997, tt2002}. If the planet mass is large enough (up to Jupiter mass), however, the planet can carve a gap near its orbit and receive a reduced torque \citep{lp1979,pl1984,cm2006}. This type of migration is termed type II, occurring more slowly compared to type I migration. 

Spiral waves are not the only source of the gravitational torque: the material in the corotation region can also exchange angular momentum with a planet during its horseshoe turn \citep{w1991}. This torque is referred to as the corotation torque and depends on the radial gradients of vortensity and entropy in the disk \citep{pp2008, bm2008}. If the gradients are steep enough, the corotation torque can halt or even reverse the inward migration driven by the Lindblad torque \citep{pp2009}, although it may vanish after a few horseshoe libration periods due to the process known as the corotation torque saturation \citep{m2001, ol2003}. This is because the corotation region is filled with various fluid elements with different libration periods, leading to a phase mixing that tends to flatten the vortensity and entropy gradients. Diffusive processes such as viscous dissipation and radiative transfer can possibly prevent the corotation torque saturation \citep{m2001}.

The migration rate is also an important factor for the corotation torque. While the material crossing the co-orbital region of a migrating planet accelerates the ongoing migration, the material trapped inside the horseshoe region slows down the migration as it is comoving with the planet. When the difference, called the co-orbital mass deficit, between the mass flowing across the co-orbital region and the mass trapped inside the horseshoe region becomes comparable to or larger than the planet mass, the migration is referred to as type III. The type III migration is important for intermediate-mass (Saturn-sized) planets that can open a partial gap \citep{mp2003} or in disks with a steep surface density gradient \citep{pa2008}.

In a disk with a vortensity gradient, a planet migration develops a vortensity contrast between the co-orbital region and the background disk. For migrating low-mass planets initiated by the Lindblad torque, this vortensity contrast is equivalent to the co-orbital mass deficit in the type III migration \citep{p2014}. The corotation torque arising from the vortensity contrast is called the dynamical corotation torque. It can be directed either radially inward or outward depending on the background vortensity gradient, affecting the type I migration of low-mass planets. 

While the aformentioned theoretical studies are useful to understand the physics behind planet migration and predict the total torque \citep{pp2009, pb2010}, they were limited to two-dimensional (2D) disks with simplified thermodynamics using a locally isothermal or adiabatic equation of state. However, real protoplanetary disks are vertically and thermally stratified, with hotter surface layers when the stellar irradiation dominates the temperature structure \citep{cg1997, lt2021} and subject to radiative diffusion. Finite disk thickness and diffusive processes can not only weaken the planet-disk interactions \citep{tt2002, mr2020} but also introduce intrinsically three-dimensional (3D) processes, such as buoyancy resonance \citep{zs2012}, which is absent in 2D disks. 

Similarly to the Lindblad resonance, the buoyancy resonance occurs in the regions where the vertical oscillation frequency of a fluid element is equal to integral multiples of $\Omega-\Omega_p$, where $\Omega$ and $\Omega_p$ stand for the angular frequency of gas and planet, respectively. Since the oscillation frequency varies with the height and $\Omega$ is a function of the radius, density ridges induced by the buoyancy resonance take a form of tilted planes \citep{zs2012, zd2015} and can exert the torque comparable to the Lindblad torque for low-mass planets \citep{zs2012, lz2014}. More recently, \citet{mn2020} found that the buoyancy resonance in a 3D adiabatic disk can change the sign of the dynamical corotation torque compared to the 2D case of \citet{p2014}, accelerating the inward migration.

However, all 3D studies mentioned above still considered a simple isothermal or adiabatic equation of state, so that they were unable to capture the effects of radiative diffusion that can potentially change the behavior of the corotation torque (e.g., \citealt{pp2008, kb2009}). For instance, it is well known that the buoyancy resonance is dependent quite sensitively on the disk structure and thermodynamics \citep{zs2012, bt2021}. Thermal diffusion is likely to dissipate the propagation of gravity waves \citep{vf2005} and inhibit the growth of temperature perturbations \citep{bt2021}, tending to suppress the buoyancy resonance. In order to properly assess the effect of the buoyancy resonance to the planet migration, therefore, it is necessary to run 3D simulations with the effect of thermal transport due to radiative diffusion included.

In this paper, we investigate the effect of the buoyancy resonance on the migration of a low-mass planet embedded in a 3D inviscid disk with radiative diffusion. This work extends the 2D simulations of \citet{p2014} by including the vertical dimension, and also the 3D simulations of \citet{mn2020} by incorporating radiative diffusion. We fix the disk parameters and planet mass, and consider four models with or without radiative diffusion and with or without planet migration. By comparing the total torque exerted on the planet from each model, we quantify the effect of the radiative diffusion on the buoyancy resonance and the dynamical corotation torque. 

To realize the thermal transport inside a disk, we adopt the method of flux-limited diffusion (FLD, \citealp{lp1981}) which calculates radiation flux based on the opacity and the radiative energy gradient. The FLD method was used in various studies including the planet accretion \citep{ab2009}, inner disk structure \citep{ff2016}, and planet migration \citep{kb2009, lc2014, bm2015} in 3D disks. For example, previous studies showed that an inclusion of the radiative diffusion induces thermal perturbations, referred to as ``cold fingers'' in the vicinity of the planet, acting as a source of additional torque \citep{kb2009, lc2014}. However, these studies considered viscous radiative disks and did not focus on the buoyancy resonance that might have been suppressed by both viscosity and radiative diffusion. The present work is complementary to these studies in that the former helps to understand how the buoyancy resonance in an inviscid disk is affected by the radiation transport.

This paper is organized as follow. In Section \ref{sec:method}, we describe our numerical methods and model parameters. In Section \ref{sec:results}, we compare the torques on a fixed or moving planet in purely adiabatic and radiative disks. Then, we explore how the radiative diffusion changes the torques exerted on the planet. In Section \ref{sec:discuss}, we discuss our results in terms of the radiative diffusion timescale, and present the results of planet migration in a polytropic disk where buoyancy resonance is absent. Finally, we conclude the present work in Section \ref{sec:conc}.

\section{Methods} \label{sec:method}

\subsection{Basic Equations} \label{subsec:equations}

We model a protoplanetary disk as a 3D, non-self-gravitating, unmagnetized, inviscid, gaseous disk around a central star with mass $M_*=1\unit{M_\odot}$. The disk has an embedded planet with mass $M_p$ initially at radius $r_p$. The planet is either forced to rotate at fixed angular frequency $\mathbf{\Omega}_0 = (GM_*/r_p^3)^{1/2}\mathbf{\hat z}$ or allowed to migrate radially inward due to instantaneous gravitational torque. 
We choose spherical polar coordinates $(r, \theta, \phi )$ in the frame rotating at $\mathbf{\Omega}_0$. The equations of hydrodynamics we solve read 
\begin{align}
\frac{\partial\rho}{\partial t} + \nabla \cdot (\rho \mathbf{v} ) &= 0, \label{eq:cont} \\
\frac{\partial\mathbf{v}}{\partial t} + (\mathbf{v}\cdot\mathbf{\nabla}) \mathbf{v} &= -\frac{\nabla P}{\rho}-\nabla\Phi- 2\mathbf{\Omega}_0 \times\mathbf{v} + \Omega_0^2\mathbf{R} , \label{eq:moment} \\
\frac{\partial e}{\partial t} + \nabla \cdot (e \mathbf{v} )  &= - P \mathbf{\nabla}\cdot\mathbf{v} +Q_{\text{rad}}. \label{eq:energy}
\end{align}
Here, $\rho$ is the gas density, $t$ is time, $\mathbf{v}=(v_r, v_\theta, v_\phi)$ is the velocity in the rotating frame, 
$P=(\gamma-1)e$ is the gas pressure with the adiabatic index $\gamma=1.4$, $\mathbf{R}$ is the cylindrical radius, and $e$ is the internal energy per unit volume. In Equation \eqref{eq:moment}, $\Phi=\Phi_* + \Phi_p$ is the total gravitational potential consisting of two parts: $\Phi_*=-GM_*/r$ is from the central star and $\Phi_p$ is due to the planet. We shall specify $\Phi_p$ in Section \ref{subsec:diskmdl}. 

In Equation \eqref{eq:energy}, $Q_{\text{rad}}$ is the heating rate per unit volume due to radiative diffusion. For an adiabatic disk, we take $Q_{\text{rad}}=0$ and consider only the compressional heating ($-P \nabla \cdot \mathbf{v}$). For a radiative disk, we follow the two-temperature approach (e.g., \citealt{bc2013}) in which $Q_{\text{rad}}$ is 
given as 
\begin{equation}
  Q_{\text{rad}} = \rho\kappa_P(cE_R - 4\sigma T^4), 
\end{equation}
where $\kappa_P$ is the Planck mean opacity, $c$ is the speed of light, $E_R$ is the local radiative energy density, $\sigma$ is the Stefan-Boltzmann constant, and $T$ is the gas temperature. Under the two-temperature FLD approximation, $E_R$ evolves as 
\begin{align}
\frac{\partial E_R}{\partial t} + \mathbf{\nabla}\cdot\mathbf{F} = -Q_{\text{rad}}, 
\end{align}
where $\mathbf{F}$ is the radiative flux given by 
\begin{equation}
\mathbf{F} = -\lambda_{\text{lim}}\frac{c}{\rho\kappa_R}\nabla E_R,  
\end{equation}
with the flux limiter $\lambda_{\text{lim}}$ and the Rosseland mean opacity $\kappa_R$. Although $\kappa_P$ and $\kappa_R$ are different from each other in reality, taking them equal is a first order approximation \citep{bc2013}: we adopt the opacity law $\kappa = \kappa_i \rho^a T^b $, with the coefficient $\kappa_i$ and the power indices $a$ and $b$ given in Table 3 of \citet{bl1994}. The flux limiter describes the transition from the optically thin to thick regimes, for which we use the form given in Equation (9) of \citet{k1989}. We have tested the radiative diffusion module described above on various problems including the oscillatory migration with accretion heating in a 3D disk with non-uniform opacity \citep{cl2019}, as presented in Appendix, and confirmed that our implementation is accurate and reliable. 

\subsection{Method}\label{subsec:gridbc}

We integrate the basic equations using the modified FARGO3D code in spherical polar coordinates \citep{m2000,bm2016}. Our computational domain extends from $r_{\text{min}}=0.8\unit{au}$ to $r_{\text{max}} = 4\unit{au}$ in the radial direction, from $\phi_{\text{min}}=-\pi$ to $\phi_{\text{max}}=\pi$ in the azimuthal direction, and $\theta_{\text{min}} = \pi/2-\arctan(3H/r)$ to $\theta_{\text{max}} = \pi/2$ in the polar (or vertical) direction, where $H$ is the disk scale height. We set up a non-uniform, logarithmically spaced grid with 805 cells in the radial direction, 3141 cells in the azimuthal direction, and 75 cells in the polar direction. The resulting grid spacing is $\Delta r\approx 0.004 \unit{au}$ at $r=r_p$, $\Delta\phi\approx0.002\unit{rad}$, and $\Delta\theta\approx0.002\unit{rad}$, corresponding to 25 zones per scale height, i.e., $H/(r\Delta \theta) = 25$.

We impose the periodic boundary conditions at the azimuthal boundaries ($\phi =\pm \pi$), while adopting the reflection boundary conditions at the midplane $(\theta =\pi/2)$ as we only model the upper hemisphere of the disk. At the upper polar boundary, we extrapolate $\rho$ and $v_{\phi}$ from the active to ghost zones using the initial disk profile (see Section \ref{subsec:diskmdl}), while adopting the continuous boundary condition for $v_r$ and the reflection boundary condition for $v_{\theta}$. For the internal energy density $e$, we fix the temperature at the upper boundary to the initial profile throughout the simulation, assuming that the disk is constantly irradiated by the central star. At the radial edges, we apply the same boundary conditions for $\rho$ and $v_{\phi}$ as in the upper polar boundary, the reflecting boundary condition for $v_r$, and the continuous boundary conditions for $v_{\theta}$, $e$, and $E_R$. Additionally, we impose the wave-damping zones at the radial boundaries to prevent the wave reflection \citep{de2006,mn2019}. 

\subsection{Model Parameters}\label{subsec:diskmdl}

Following \citet{mn2020}, we initially consider a disk with temperature $T = 306.6 (r/1\unit{au})^{-1}\unit{K}$, surface density $\Sigma = 3.8\times10^{-4}(r/1\unit{au})^{-0.5}\,\unit{M_{\odot}}\ \unit{au^{-2}}$, and mean molecular weight $\mu = 2.3\unit{g}\ \unit{mol^{-1}}$. Assuming that the disk is initially isothermal in the vertical direction and has a constant aspect ratio $h\equiv H/r=0.05$, the corresponding density profile is 
\begin{equation}
  \rho(r,\theta) = \rho_0(\sin\theta)^{-2.5+h^{-2}} \left(\frac{r}{1\unit{au}}\right)^{-1.5},
\end{equation}
where $\rho_0=3.032\times10^{-3} \,\unit{M_{\odot}} \unit{au^{-3}}$ is the equatorial density at $r=1\,\unit{au}$
\citep{mb2016}. We initially take $v_r=v_\theta=0$, and set $v_{\phi}$ to the values that ensure an exact balance between the centrifugal force and the pressure gradient force.

We consider a planet with mass $M_p = 6.67\unit{M_{\oplus}}$, corresponding to the mass ratio $q \equiv M_p/M_* = 2\times 10^{-5}$, and place it at $r_p = 2\unit{au}$ initially. Similarly to \citet{mn2020}, we treat the planet as a uniform density sphere with radius $\epsilon = 0.1 H(r_p)$. The gravitational potential of the planet is then 
\begin{equation}
\Phi_p =
  \begin{cases}
  -{GM_p}/{d}, & \text{for}\; d>\epsilon, \\
  -{GM_p(3\epsilon^2-d^2)}/{(2d^3)}, & \text{for}\; d\leq\epsilon,
  \end{cases}
\end{equation}
where $d\equiv |\mathbf{r}-\mathbf{r}_p|$, with $\mathbf{r}_p(t)$ being the position vector of the planet at time $t$.

\begin{figure*}[t!]
\epsscale{1.0}
\plotone{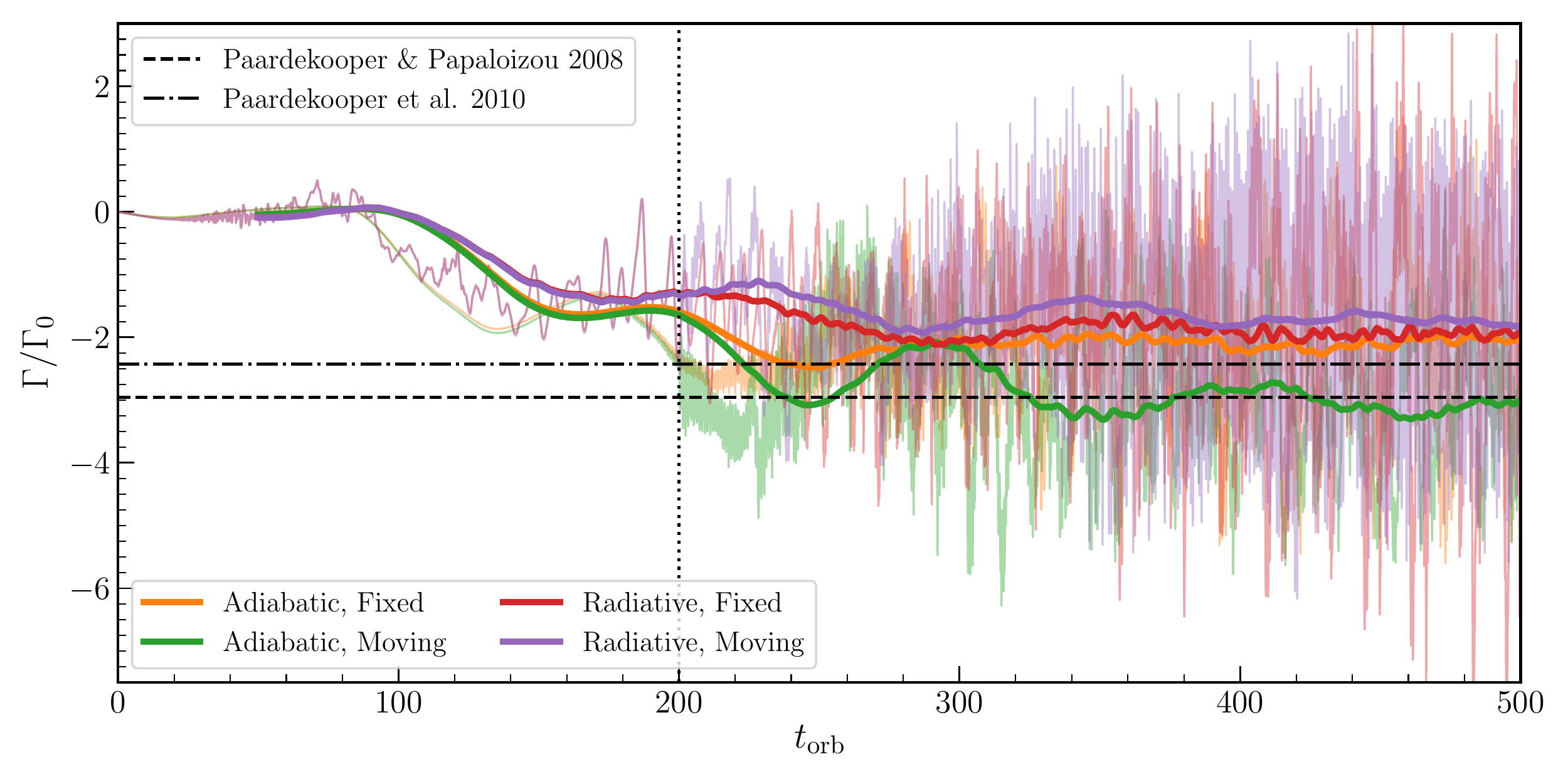}
\caption{Temporal variations of the normalized torques on the fixed and moving planets. Light lines give the instantaneous torque, while the solid lines correspond to the mean values averaged over 50 orbits. The dashed and the dot-dashed lines correspond to the Lindblad torque \citep{pp2008} and the Lindblad torque combined with the linear corotation torque \citep{pb2010} in a corresponding 2D disk, respectively. In the 3D purely adiabatic disk, the torque on the fixed planet (\textit{orange}) is closer to the combined torque, while the torque on the moving planet (\textit{green}) converges to the Lindblad torque, indicating that the dynamical corotation torque strengthens the inward migration. In the 3D radiative disk, however, the sign of the dynamical corotation torque is opposite to make the torque on the fixed planet (\textit{red}) slightly more negative than the torque on the moving planet (\textit{purple}). \label{fig:tqwk}}
\end{figure*}


To avoid transients in the flows caused by a sudden introduction of the planet, we slowly introduce the planet at $\mathbf{r}_p(0)=(r_p, \pi/2, 0)$ and increase its mass linearly with time from zero to $M_p$ over $200t_\text{orb}$, where $t_\text{orb} \equiv 2\pi/\Omega_0= 2.83\unit{yr}$ is the orbital time of the planet. For models without planet migration, we keep the planet position unchanged for next $300t_\text{orb}$. For models with planet migration included, however, the planet is allowed to move due to the instantaneous gravitational torque it receives. We note that the planet is subject to the combined gravitational potentials of the central star and the disk, while the disk is governed by the stellar potential alone. The mismatch between the potentials on the planet and disk can shift the resonances, overestimating the torque on the planet \citep{bm2008}. In order to alleviate this issue, we adopt the workaround given in FARGO3D that considers only the non-axisymmetric part $\delta \rho \equiv \rho - \left< \rho\right> $ of the disk density when calculating the torque on the planet. Here, the angle brackets denote the azimuthal average, i.e., $\left<\rho\right > \equiv \int \rho d\phi/(2\pi)$.

In addition, the regions inside the Hill sphere of the planet is poorly resolved and often exhibit temporal fluctuations in the gas density. To remove the noisy contributions of the Hill sphere, we calculate the torque as 
\begin{equation}
  \Gamma = GM_p \iiint \frac{\delta \rho(\mathbf{r})\xi(d)}{|\mathbf{r}-\mathbf{r}_p|^3}
  \mathbf{r}_p\cdot (\mathbf{r}-\mathbf{r}_p)
  r^2\sin\theta drd\theta d\phi,
\end{equation}
where $\xi$ is the tapering function defined as
\begin{equation}
\xi(d) =
	\begin{cases}
    0, & \text{for} \;d/r_H < 1/2, \\
    \sin^2(d/r_H-1/2), & \text{for} \;1/2\leq d/r_H\leq 1,\\
    1, & \text{otherwise}, 
	\end{cases}
	\label{eq:hillcut}
\end{equation}
where $r_H\equiv (q/3)^{1/3}r_p$ is the Hill radius
\citep{mn2020}.

For a radiative disk, we initialize the radiative energy density as $E_R = 4\sigma T^4/c$. This makes the disk slightly out of equilibrium. We thus relax the initial disk described above for $50t_\text{orb}$ in the $r$--$\theta$ plane. The resulting equilibrium configuration differs from the initial one only by less than $3\%$ in $\rho$ and $T$.

\section{Results}\label{sec:results}

In this section, we will first show that the results of the adiabatic models agree with those from \citet{mn2020} in terms of the torque histories. We will then present the results of the radiative models.

\subsection{Adiabatic Disk}\label{subsec:migration}

Figure \ref{fig:tqwk} plots the torque exerted on the fixed planet (orange) and on the moving planet (green) embedded in the adiabatic disk: the thin and thick lines correspond to the instantaneous and time-averaged values over $50t_\text{orb}$, respectively. The torque is normalized by $\Gamma_0\equiv(q/h)^2\Sigma_pr_p^4\Omega_p^2$, where $\Sigma_p$ is the disk surface density at the planet position \citep{tt2002}. While the instantaneous torque exhibits rapid fluctuations due to vortices generated in the corotation region, the time-averaged torque undergoes small-amplitude oscillations and converges slowly to a quasi-steady value $\Gamma/\Gamma_0 \approx -2.01$ for the fixed planet and $\Gamma/\Gamma_0 \approx -2.99$ for the moving planet. 

\begin{figure*}[t!]
\centering
\epsscale{1.0}
\plotone{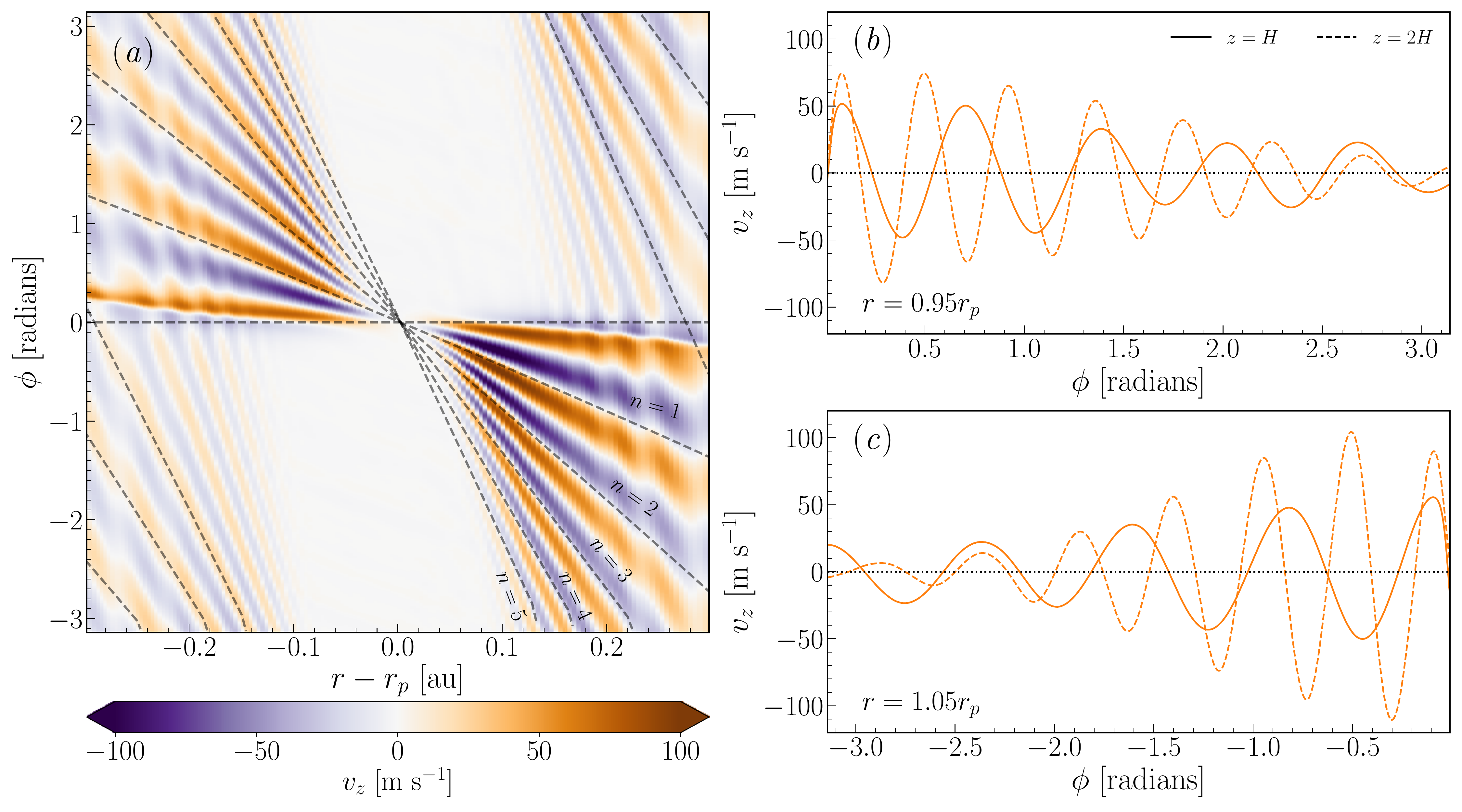}
\caption{Vertical velocity perturbations $v_z$ of the buoyancy resonance near the planet averaged over $t=200$--$500t_\text{orb}$ from the adiabatic disk model with the fixed planet (\textit{a}) in the $r$--$\phi$ planet at $z/H=2$, (\textit{b}) along $\phi$ at $r=0.95r_p$, and (\text{c}) along $\phi$ at $r=1.05r_p$. The dashed lines in (\textit{a}) draw the predicted resonance positions (Equation \ref{eq:buoy} for $n=1,\cdots,5$), which are in good agreement with the numerical results.
\label{fig:buoy}}
\end{figure*}

For comparison, Figure \ref{fig:tqwk} also plots the 2D Lindblad torque $\Gamma/\Gamma_0\approx-2.96$ (\citealt{pp2008}; dashed line) as well as the Lindblad torque combined with the linear corotation torque $\Gamma/\Gamma_0\approx-2.42$ (\citealt{pb2010}; dot-dashed line), assuming that the disk is infinitesimally thin. It is interesting to note that the torque on the fixed plant in our 3D simulation is close to the Lindblad torque plus the linear corotation torque in the corresponding 2D disk. Although this appears to contradict the previous prediction that the corotation torque is saturated in the 2D inviscid disk, the dynamics in the 3D disk may differ significantly from the 2D counterpart to change the behavior of the corotation torque (see also \citealt{fa2015,jm2017}).

A migrating planet additionally receives the dynamical corotation torque $\Gamma_\text{hs}$. \citet{p2014} showed that $\Gamma_\text{hs}$ in a 2D thin disk amounts to 
\begin{equation}\label{eq:dyntq}
\Gamma_\text{hs} = 2\pi\left[1-\frac{\omega_0(r_p)}{\omega_c}\right]\Sigma_pr_p^2x_s\Omega_p\frac{dr_p}{dt},
\end{equation}
where $\omega_0(r_p)$ is the vortensity of the unperturbed disk at the planet position, and $\omega_c$ and $x_s$ are the characteristic vortensity and the half-width of the horseshoe region, respectively\footnote{In a 2D thin disk rotating at angular frequency $\mathbf{\Omega}_0 \hat{\mathbf{z}}$, $\omega=|\nabla \times \mathbf{v}+2\mathbf{\Omega}_0|/\Sigma$, where $\mathbf{v}$ is the velocity in the rotating frame.}. Since the vortensity of our initial disk decreases radially outwards as $\propto r^{-1}$, $\Gamma_\text{hs}$ would be positive for a planet migrating inward (i.e., $dr_p/dt<0$) if $\omega_0(r_p)>\omega_c$. However, Figure \ref{fig:tqwk} shows that the torque on the inwardly migrating planet in the 3D disk is more negative than that on the fixed planet. This implies that the dynamical corotation torque in the 3D disk is negative.


\citet{mn2020} argued that the negative dynamical corotation torque inside the horseshoe region is caused by the buoyancy resonance which is absent in 2D disks. In a 3D disk, internal gravity waves or buoyancy waves are characterized by the Brunt-V\"{a}is\"{a}l\"{a} frequency 
\begin{equation}
N(z) = \sqrt{\frac{g_z}{\gamma}\frac{\partial}{\partial z}\left[\ln\left(\frac{P}{\rho^\gamma}\right)\right]},
\label{eq:bv}
\end{equation}
with the vertical gravity $g_z=-d\Phi_*/dz$. The buoyancy waves experience the gravitational force of the embedded planet as they propagate vertically, and grow in amplitudes at the positions where $N(z)=\pm m(\Omega_p-\Omega)$, where the integer $m$ denotes the azimuthal wavenumber. This buoyancy resonance is analogous to the Lindblad resonance that excites and amplifies density waves at the locations where the epicycle frequency is equal to $\pm m(\Omega_p-\Omega)$ \citep{zs2012}.
For waves with the azimuthal wavelength $\lambda =2\pi/m$, the resonance position can be calculated as a function of $r$ and $z$, Then, the line of constant phase $2n\pi$ (for integer $n$) occurs at $\phi=n\lambda$, or 
\begin{equation}
\phi = \frac{\pm 2n\pi}{(1-1/\gamma)^{1/2}} \frac{(\Omega_p-\Omega)H}{\Omega_K(R)z}\left(1+\frac{z^2}{R^2}\right)^{3/2},
\label{eq:buoy}
\end{equation}
where $R=(r^2+z^2)^{1/2}$ \citep{zd2015}.

Figure \ref{fig:buoy}($a$) plots the vertical velocity perturbations $v_z$ created by the buoyancy resonance in the $r$--$\phi$ plane averaged over $t=200$--$500t_\text{orb}$ at $z=2H$ obtained from the fixed-planet, adiabatic-disk model. Also plotted as the dashed lines are Equation \eqref{eq:buoy} for $n=1,\cdots, 5$,
which well trace the resonance positions in the simulation. Figure \ref{fig:buoy}($b,c$) plots the distributions of $v_z$ along the azimuthal direction at $z/H=1,2$ and $r/r_p=0.95, 1.05$, showing that the buoyancy resonance is stronger in the regions closer to the plant and at higher-altitude regions. Note that the strongest  perturbations lying close to the $\phi=0$ line seen in Figure \ref{fig:buoy} are not due to the buoyancy resonance but parts of the spirals produced by the Lindblad resonance.

\begin{figure}[t!]
\epsscale{1.2}
\plotone{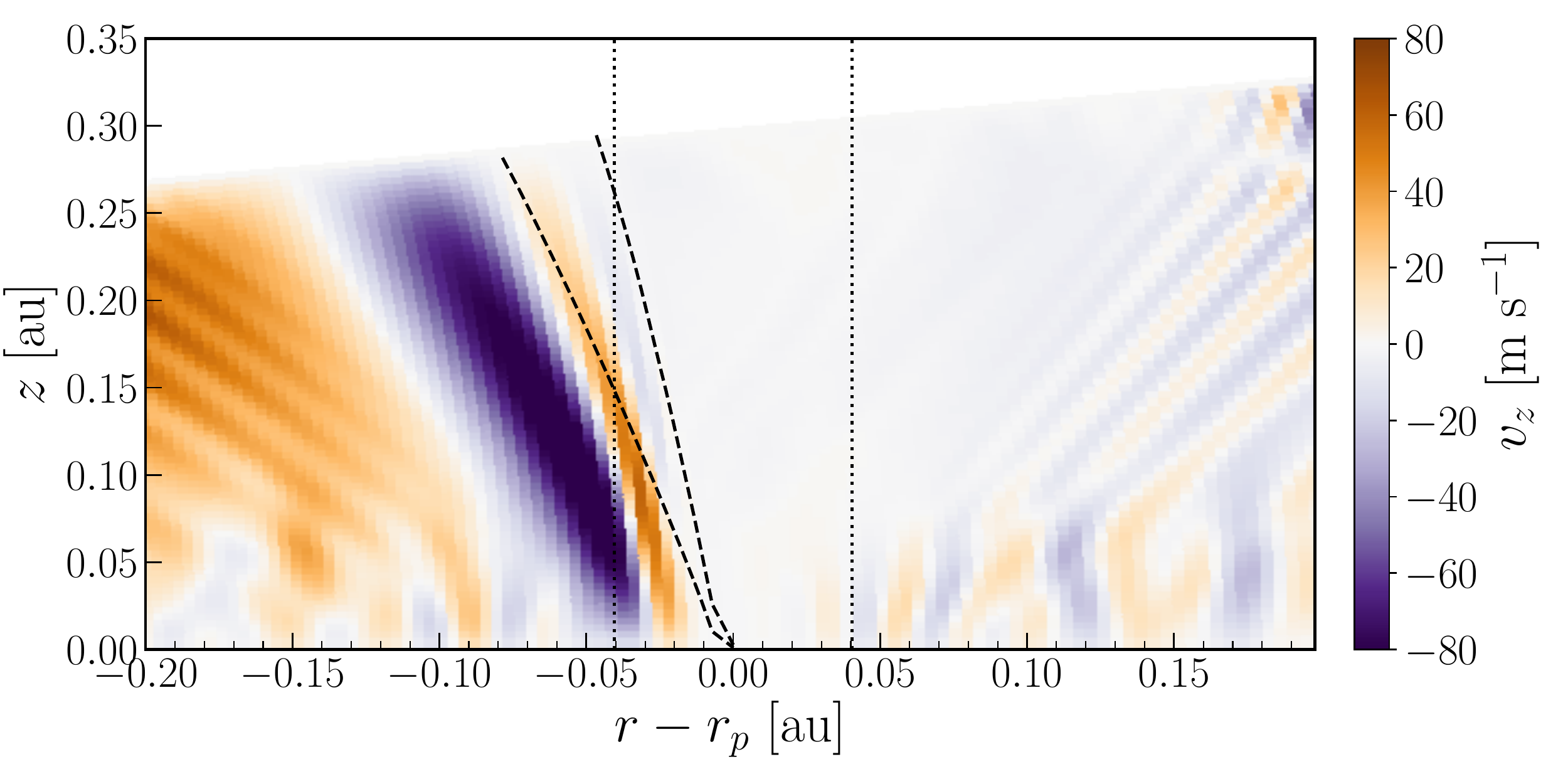}
\caption{Vertical velocity perturbations $v_z$ in the $r$--$z$ plane at $\phi=0.2$ rad from the adiabatic disk model with a fixed planet averaged over $t = 200-500t_{\mathrm{orb}}$. The dashed lines draw Equation \eqref{eq:buoy} for $n=1$ and $2$.
The vertical dotted lines mark the edges of the the corotation regions, which are affected by the buoyancy resonance at low $|z|$. 
} \label{fig:buoycorot}
\end{figure}

Figure \ref{fig:buoy} gives an impression that the buoyancy resonance at $|z|/H=2$ is rather weak inside the horseshoe regions bounded by $|r-r_p|=0.02r_p$. To explore the vertical dependence of the buoyancy resonance, Figure \ref{fig:buoycorot} plots the velocity perturbations in the $r$--$z$ plane from the adiabatic, fixed-planet model. Note that the regions with strong $|v_z|$ move toward the planet as $|z|$ decreases, affecting the corotation regions whose boundaries are drawn by the vertical dotted lines. We also note that the predicted resonance positions do not match well with the simulation results at low-$|z|$ regions (see also \citealt{zs2012,zd2015,mn2020}). This may be because the Brunt-V\"{a}is\"{a}l\"{a} frequency becomes vanishingly small as $|z|\rightarrow0$, in which case the resonance positions may be compromised by sound waves and/or epicycle motions.

\subsection{Radiative Disk}\label{sec:rad}

\begin{figure}[t!]
\epsscale{1.2}
\plotone{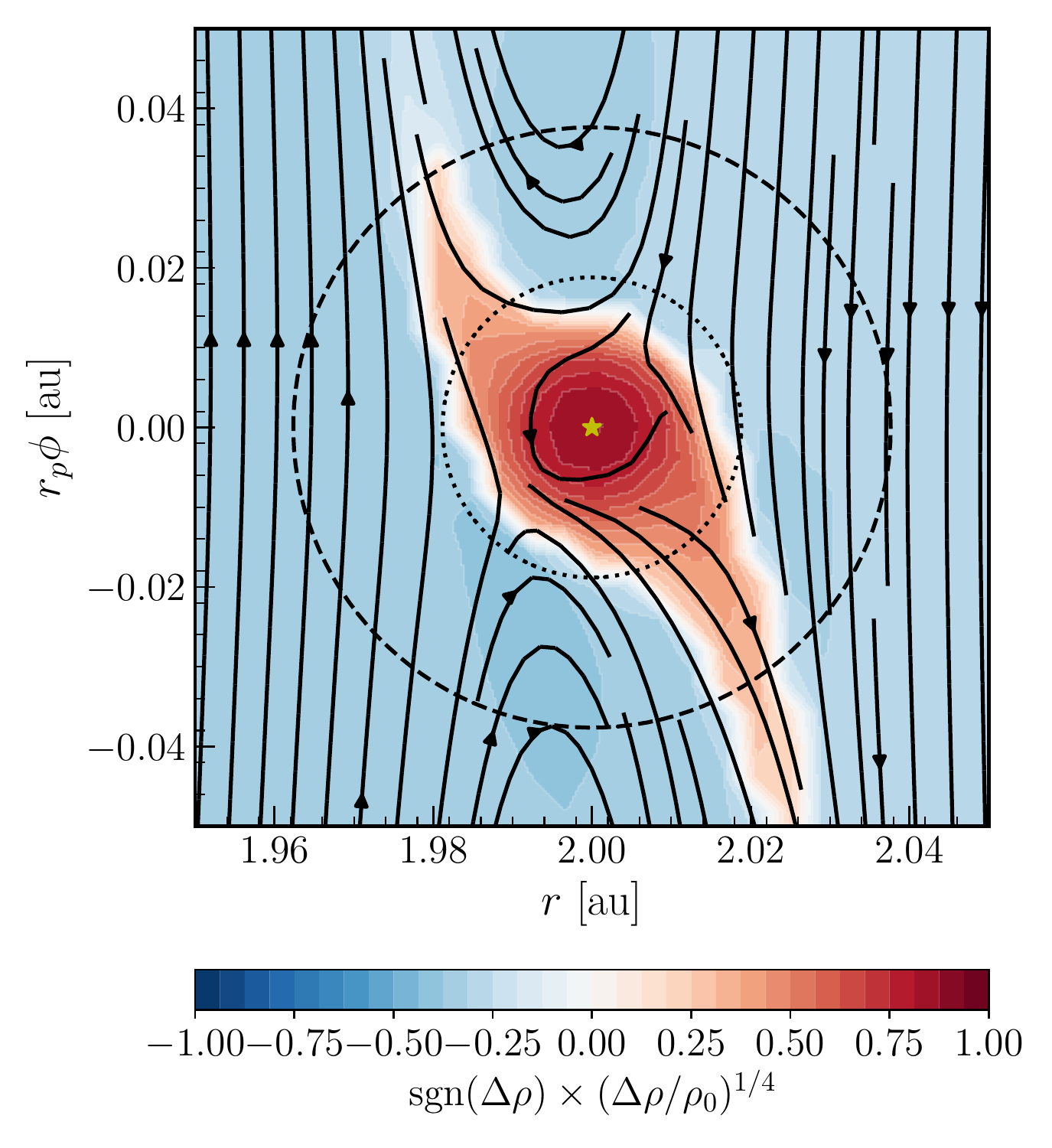}
\caption{
Difference in the gas density at $z=0$ between the adiabatic and radiative models averaged over $t=200$-$500t_\text{orb}$.
The radiative diffusion not only creates the dense regions near the planet but also two dense lobes called cold fingers \citep{lc2014}. Solid lines with arrows draw the associated streamlines, while the dashed and dotted circles mark the Hill and half-Hill spheres, respectively.} \label{fig:dens}
\end{figure}

Now we present the results of the simulations with the radiative disk. The temporal changes of the torques on the fixed and moving planets are plotted in Figure \ref{fig:tqwk} as red and purple lines, respectively. Again, light lines correspond to the instantaneous torques, while the thick lines give the values averaged over $50t_\text{orb}$: the total torque at the end of the simulation is $\Gamma/\Gamma_0 \approx -1.93$ on the fixed planet and $\Gamma/\Gamma_0 \approx -1.82$ on the moving planet.

Let us first focus on the torque acting on the fixed planet in the radiative disk. \citet{lc2014} showed that in addition to the Lindblad and corotation toruqes, 
radiative diffusion introduces thermal perturbations in the vicinity of the planet known as ``cold fingers". In the absence of radiative diffusion, the gas on horseshoe orbits is heated by compression as it approaches the planet, and cools down by rarefaction as it moves away from the planet. The radiative diffusion takes away the internal energy from the compressed region, which in turn makes the region compressed further compared to the adiabatic case. This leaves elongated structures, namely cold fingers, near the planet after the U-turns of the streamlines. The cold fingers are apparent in Figure \ref{fig:dens} which plots the difference of the gas density at $z=0$ near the planet between the adiabatic and radiative models  with a fixed planet.

\begin{figure*}[t]
\centering
\epsscale{1.0}
\plotone{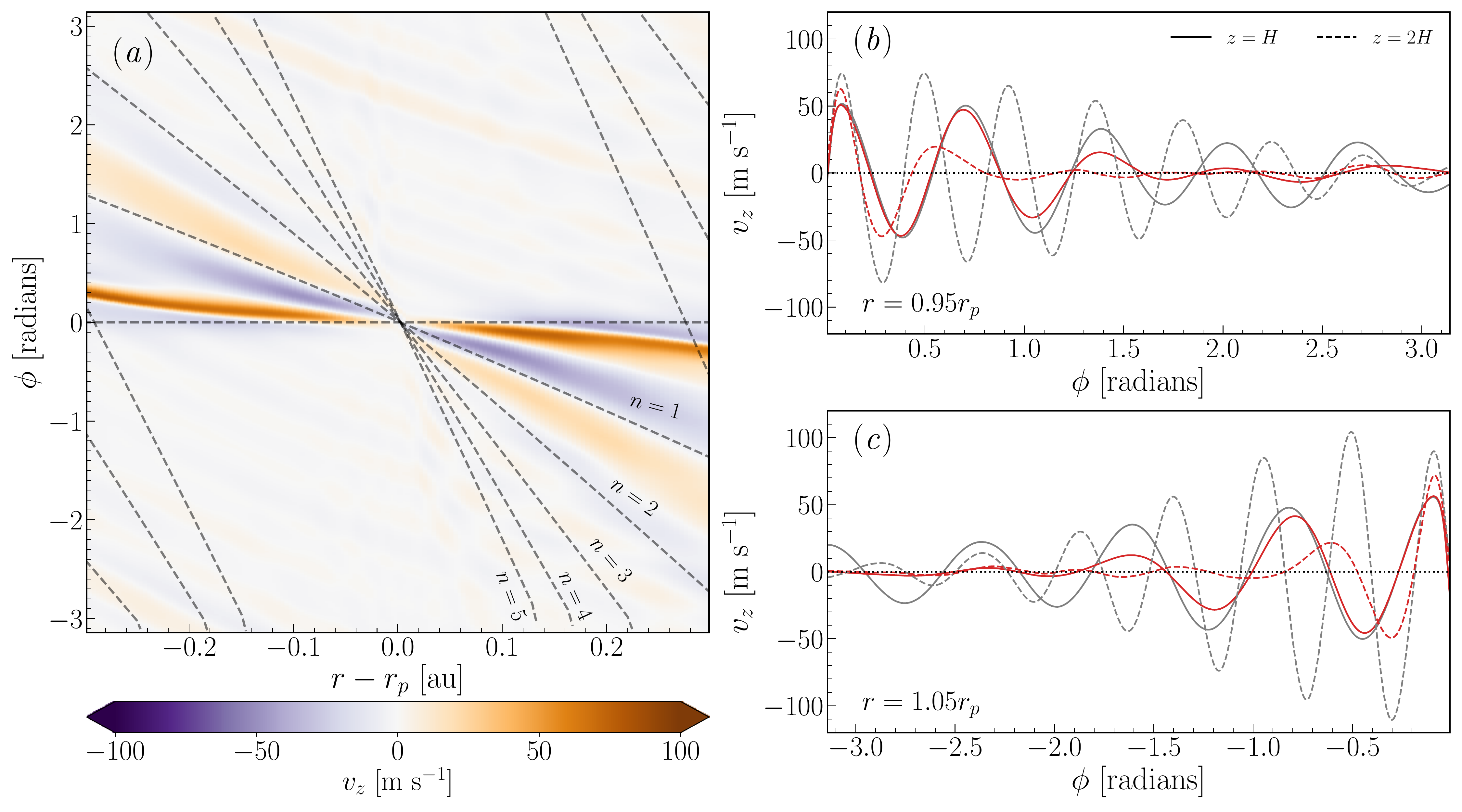}
\caption{Same as Figure \ref{fig:buoy} but for the radiative disk. For comparison, the velocity perturbations in the adiabatic disk are plotted as grey dotted lines in ($b$) and ($c$), showing that the buoyancy resonance is significantly weakened, more strongly at higher $|z|$, compared to the adiabatic counterpart. \label{fig:buoy2}}
\end{figure*}

The asymmetry of the cold fingers may contribute to the total torque on the planet: the torque from the cold fingers would be negative if the inner finger at $r<r_p$ is stronger than the outer one at $r>r_p$. This asymmetry is dependent heavily on the disk parameters as they affect gas flows around the planet \citep{lc2014}. However, when we employ the Hill cut using the tapering function (Equation \ref{eq:hillcut}), which excludes the contribution of the half-Hill sphere marked by the dotted circle in Figure \ref{fig:dens}, the impact of the cold fingers on migration becomes only moderate. The outer finger (at $r>r_p$) outside the half-Hill sphere is a bit stronger than the inner one (at $r<r_p$) in our simulations, explaining why the torque exerted on the fixed planet in the radiative disk is slightly less negative compared to the adiabatic counterpart. Nonetheless, the radiative diffusion does not cause significant changes on the torque exerted on the fixed planet by the cold fingers.

\begin{figure}[t!]
\epsscale{1.2}
\plotone{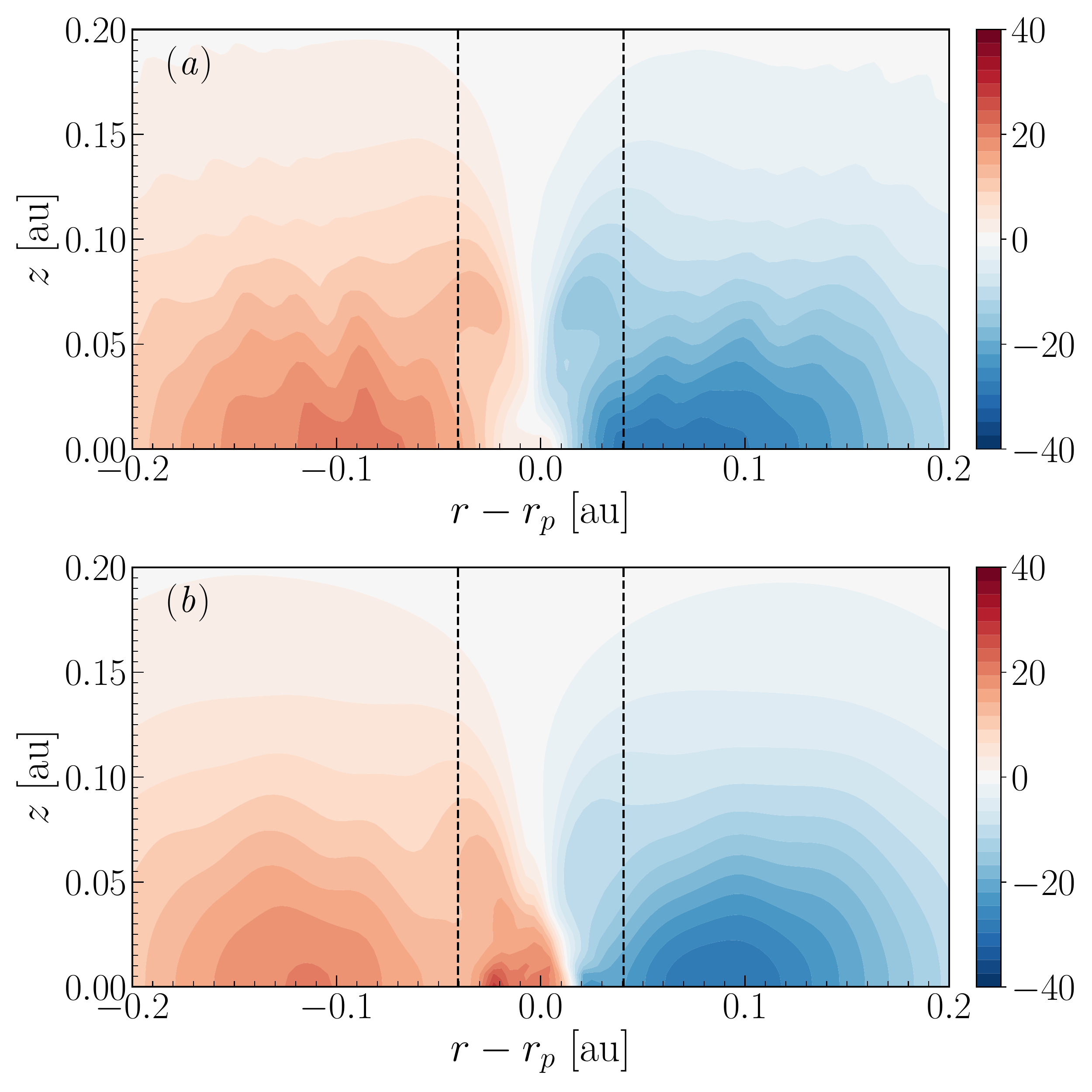}
\caption{Distributions of the azimuthally-averaged torque density
$\langle d^3\Gamma/(r^2\sin\theta drd\theta d\phi) \rangle$ 
in the $r$--$z$ plane for the fixed planet in the ($a$) adiabatic and ($b$) radiative disks. The dashed lines draw the edges of the corotation regions.} The vertical striations of the torque density in the adiabatic disk are due to the the buoyancy resonance, which are much weaker in the radiative disk. \label{fig:tqdens}
\end{figure}

On the other hand, the radiative diffusion shifts the migration behavior greatly when the planet is allowed to migrate. Figure \ref{fig:tqwk} shows that in the radiative disk the torque on the moving planet is slightly less negative than that on the fixed planet, while the former is more negative than the latter in the adiabatic disk. It appears that the dynamical corotation torque is almost absent in the radiative disk, making the moving planet receive about 40\% less torque compared to the adiabatic case.

To understand the reduced dynamical corotation torque in the radiative disk,
Figure \ref{fig:buoy2} plots the time-averaged velocity perturbations $v_z$ driven by the buoyancy resonance at $z=2H$ as well as $v_z$ at $r/r_p=0.95,1.05$ and $z/H=1,2$ from the model with the moving planet in the radiative disk. It can be readily seen that the radiative diffusion weakens the buoyancy resonance significantly compared to the adiabatic case shown in Figure \ref{fig:buoy}.

Figure \ref{fig:tqdens} compares the azimuthally-averaged torque density $\langle d^3\Gamma/(r^2\sin\theta drd\theta d\phi) \rangle$ between the adiabatic and radiative disks.
The torque density map is dominated by two lobes corresponding to the positive (at $r<r_p$) and negative (at $r>r_p$) Lindblad torques. The corrugations of the iso-torque contours reflect the effect of the buoyancy resonance \citep{mn2020}. 
Note that one of the corrugations resides inside the corotation region at $z\approx0.07\,\rm au$, indicating that the buoyancy resonance affects the corotation torque exerted on the planet.  Weaker corrugations in the radiative disk  suggest that the radiative diffusion suppresses buoyancy resonance. In addition, the cortotation torque that is saturated in the adiabatic disk becomes unsaturated in the radiative disk (e.g., \citealp{bm2008, mc2010}).

\begin{figure}[t!]
\epsscale{1.2}
\plotone{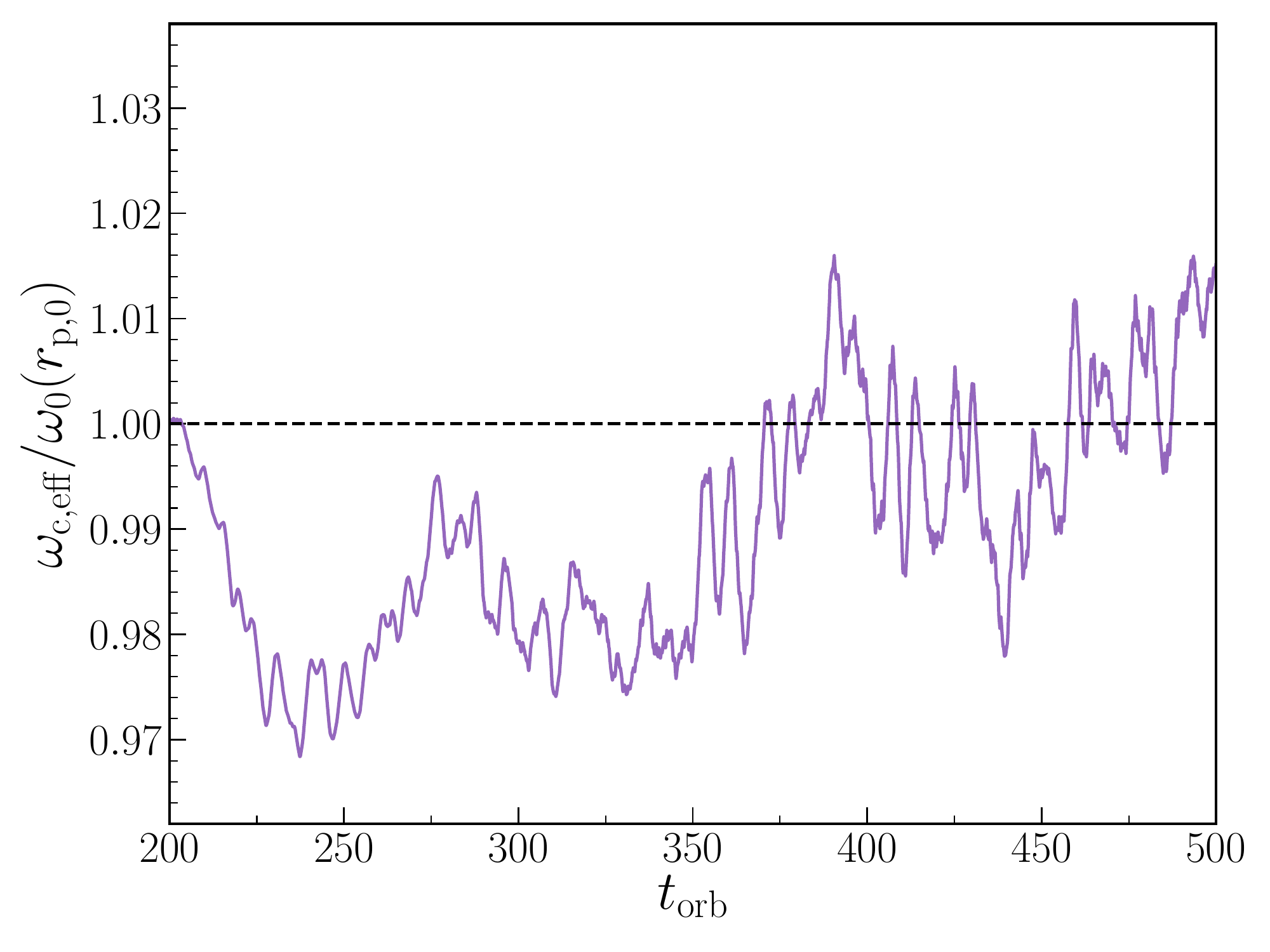}
\caption{The characteristic vortensity of a radiative disk $\omega_{\text{c,eff}}$ calculated from Equation \eqref{eq:dyntq}. Since the difference between $\omega_{\text{c,eff}}$ and $\omega_0(r_{\text{p,0}})$ is less than 3\%, we can assume that the dynamical corotation torque on the radiative disk is similar to the 2D case reported on \citet{p2014}. \label{fig:wc}}
\end{figure}

That the radiative diffusion weakens the buoyancy resonance implies that vertical gas motions in the radiative disk are reduced significantly compared to those in the adiabatic disk. In this situation, one can expect the dynamical corotation torque on a migrating planet in the radiative disk is driven by the vortensity gradient, as in 2D disks \citep{p2014, mn2020}. Assuming that Equation \eqref{eq:dyntq} is responsible for the difference in the torques on the moving and fixed planets in the radiative disk, one can calculate the effective characteristic vortensity $\omega_{\text{c,eff}}$. Figure \ref{fig:wc} plots $\omega_{\text{c,eff}}$ relative to  $\omega_0(r_{\text{p,0}})$, the vortensity at the initial planet position in the unperturbed disk, as a function of time. Note that $\omega_{\text{c,eff}}$ is within 3\% of $\omega_0(r_{\text{p,0}})$ throughout the simulation, which is consistent with  \citet{p2014} who argued that $\omega_{\text{c}}=\omega_0(r_{\text{p,0}})$ when the corotation region migrates with a planet in an inviscid 2D disk.
This supports the picture that the positive dynamical corotation torque in our radiative disk is most likely caused by the vortensity gradient of the disk.

 

\section{Discussion}\label{sec:discuss}


We first discuss the effect of the radiative diffusion on the buoyancy resonance. If a disk is optically thin, the gas will cool down by emitting thermal photons that escape the disk easily.
If a disk is instead optically thick, as in our models, thermal photons are absorbed and emitted multiple times before escaping the disk. In the latter case, the thermal transport timescale is comparable to the diffusion timescale in the vertical direction
\begin{equation}
t_d \equiv \frac{H^2}{\mathcal{D}} \approx \frac{3c_V \kappa\rho^2 H^2}{16\sigma T^3},
\end{equation}
where $\mathcal{D}\equiv16\sigma T^3/(3c_V\kappa\rho^2)$ is the diffusion coefficient and $c_V$ is the specific heat of the gas \citep{mk2017,bt2021}.  If $t_d$ is shorter than the buoyancy timescale $t_b\equiv 1/N(z)$, one can expect that the buoyancy resonance would be erased by the radiative diffusion.  Since $t_d\propto \rho^2$ in our vertically-isothermal disks
decrease faster than $t_b\propto z^{-1}(1+z^2/R^2)^{3/2}$ with increasing height, this happens in the regions with high $|z|$. The critical thermal diffusion coefficient is given by 
\begin{equation}
    \mathcal{D}_{\text{crit}} = \left(\frac{\gamma-1}{\gamma}\right)^{1/2}\Omega_KHz\left(1+\frac{z^2}{R^2}\right)^{-3/2},
\end{equation}
above which the buoyancy resonance is suppressed: 
$\mathcal{D_{\text{crit}}}\sim 8.5\times 10^{16}\rm\,cm^2\,s^{-1}$ at $r=r_p$ and $z=H$.

\begin{figure}[t!]
\centering
\epsscale{1.2}
\plotone{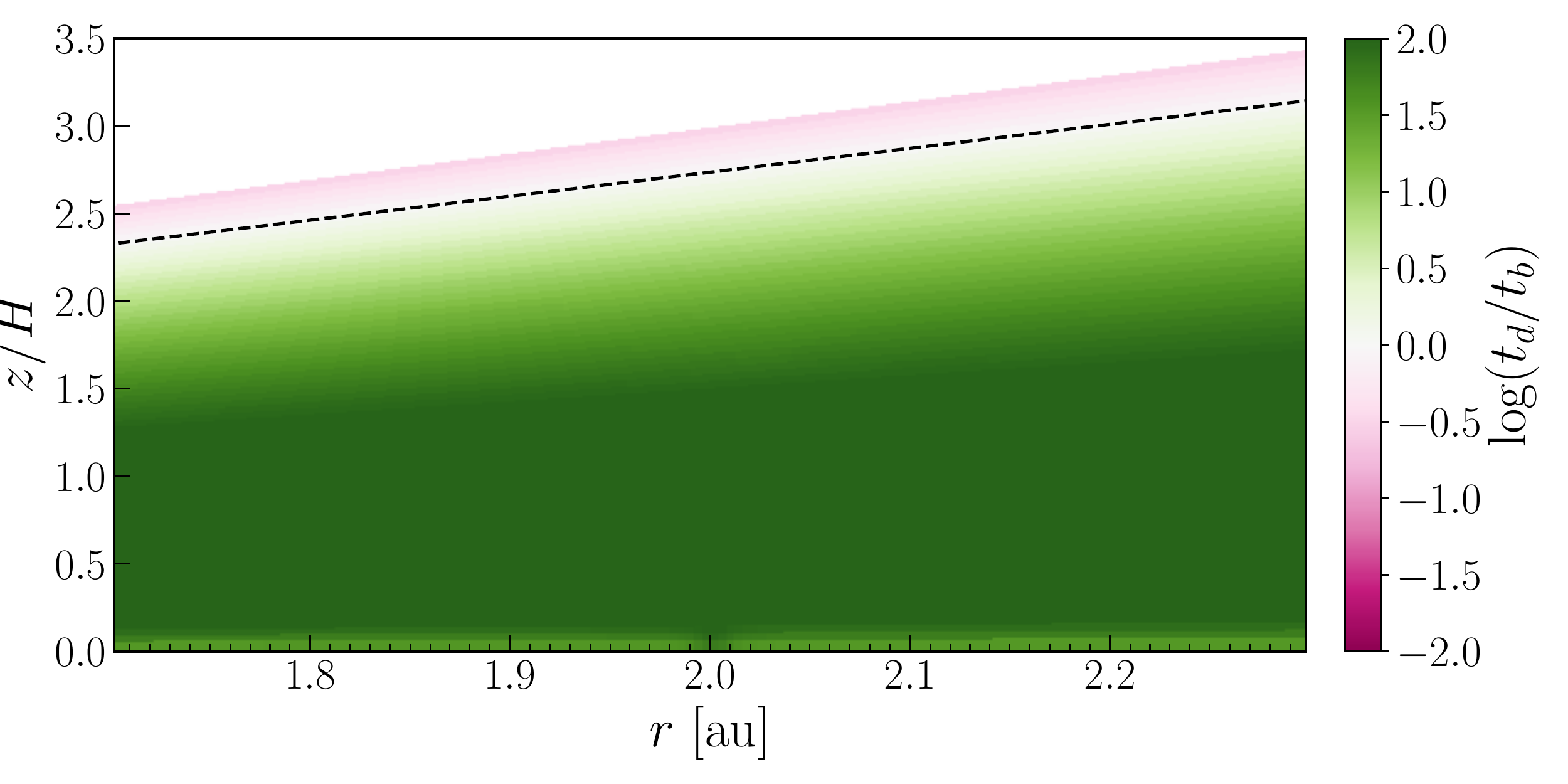}
\caption{Ratio of the radiative diffusion timescale $t_d=H^2/\mathcal{D}$ to the buoyancy response timescale $t_b=1/N$. The dashed line marks the locus with $t_d/t_b=1$, above which the buoyancy resonance is suppressed significantly.} \label{fig:cool}
\end{figure}

Figure \ref{fig:cool} plots the distribution of $t_d/t_b$ in the $r$--$z$ plane in our model disk. Clearly, $t_d/t_b\leq 1$ at $z/H\geq 2$. This suggests that the suppression of the buoyancy resonance by the radiative diffusion is considerable in a thin layer with $z\geq 2H$. At $z/H=1$, however, $t_d/t_b\sim 220$ so that the radiative diffusion is too slow to affect the buoyancy resonance. This is entirely consistent with Figure \ref{fig:buoy2} in that the radiative diffusion weakens the buoyancy resonance more strongly at higher $|z|$: only the primary mode with $n=1$ is visible at $z/H=2$, while modes with higher $n$ are erased almost completely. 

\begin{figure}[t!]
\epsscale{1.2}
\plotone{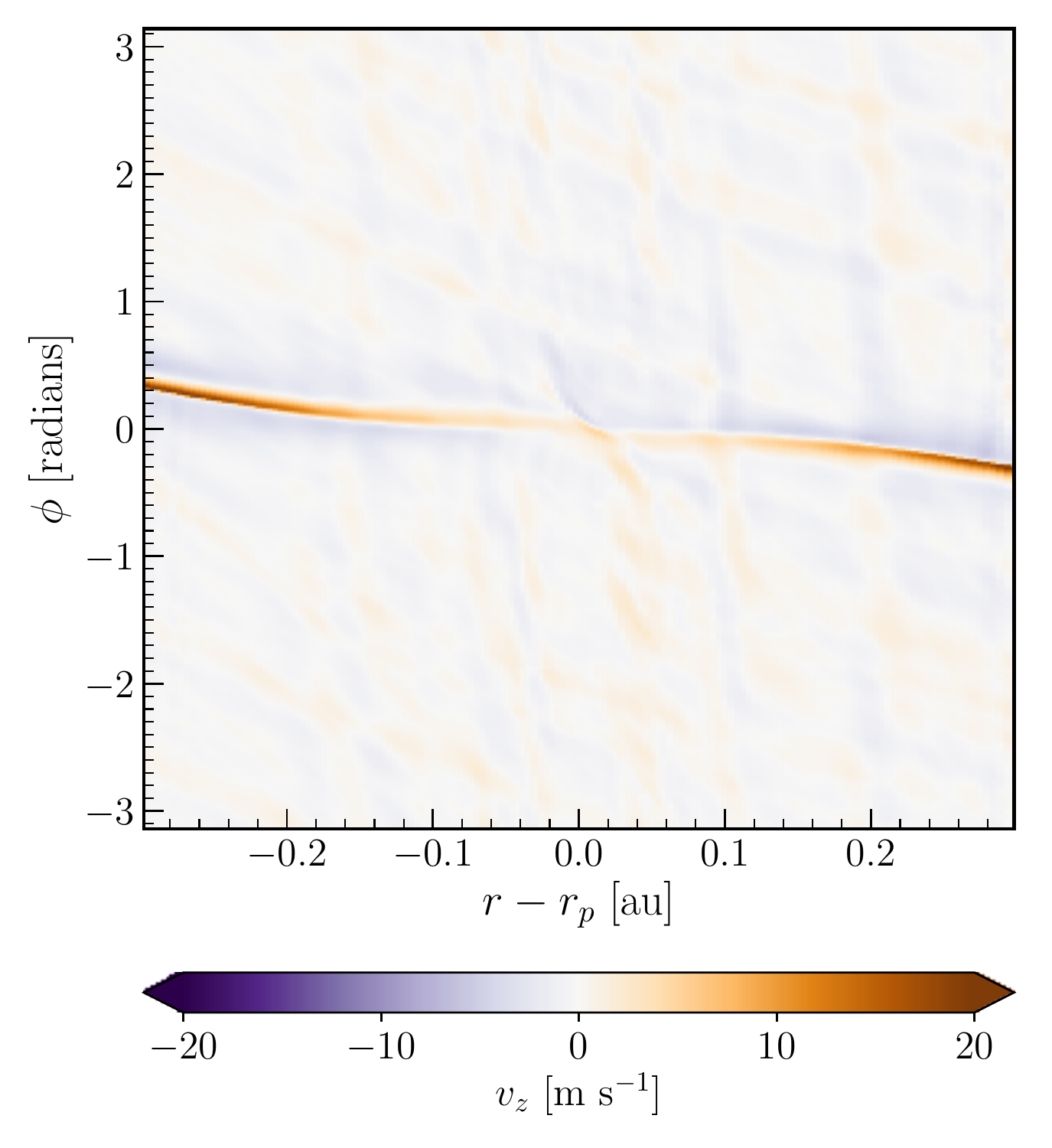}
\caption{Velocity perturbations $v_z$ in the $r$--$\phi$ plane at $z/H=2$ from the polytropic disk model with a fixed planet when $t=500t_\text{orb}$. Other than the spirals produced by the Lindblad resonance, the perturbations are almost featureless, indicating that the buoyancy resonance is absent. \label{fig:buoypoly}}
\end{figure}

\begin{figure}[t!]
\epsscale{1.2}
\plotone{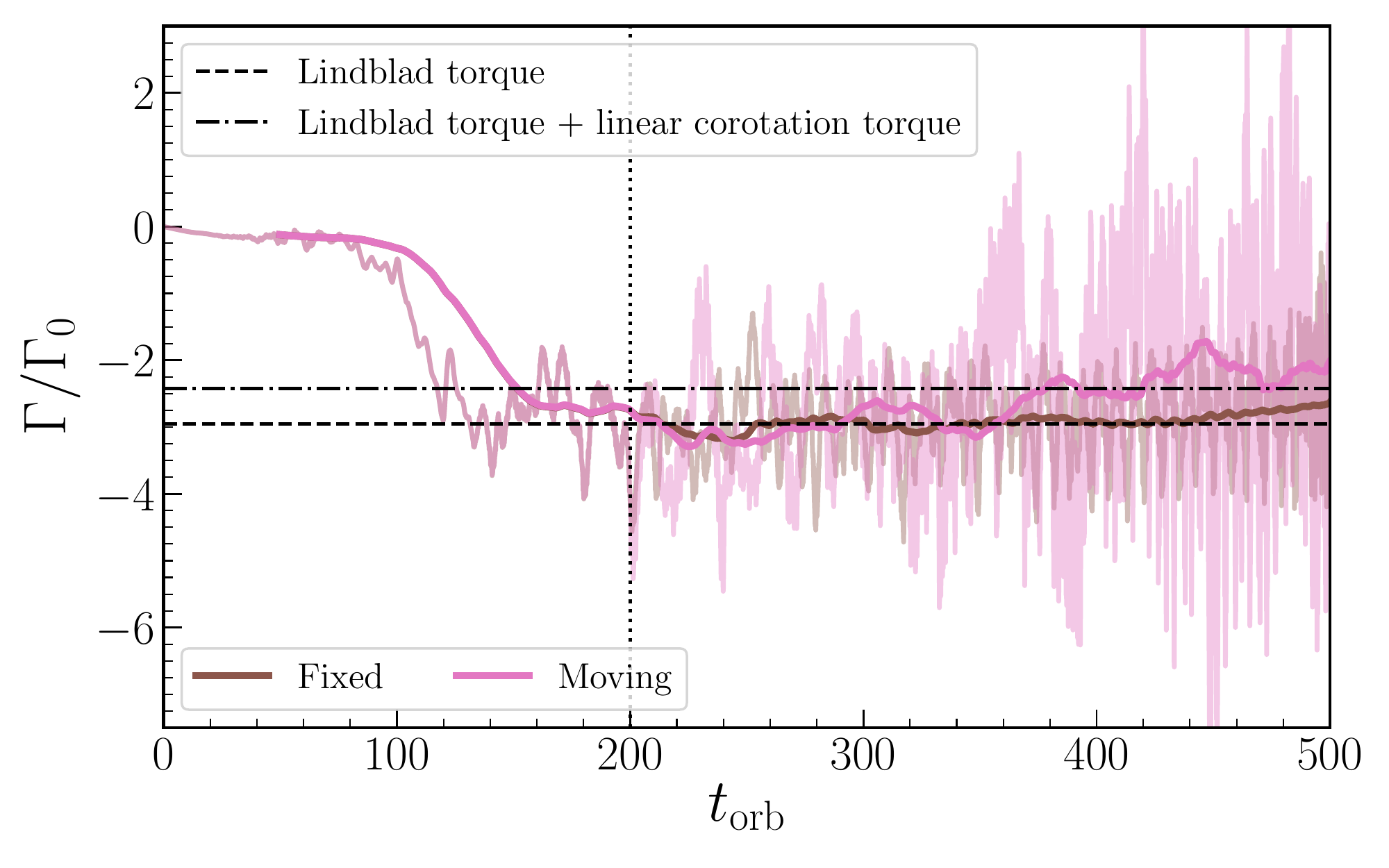}
\caption{Same as Figure \ref{fig:tqwk} but for the polytropic disk without radiative diffusion. The torque on the fixed planet is more negative than that on the moving planet, suggesting that the dynamical corotation torque is positive, which is opposite to the case in the isothermal disk.
\label{fig:plytqwk}}
\end{figure}


In Section \ref{sec:rad}, we argue that the radiative diffusion negates the negative dynamical corotation torque arising from the buoyancy resonance. To see if the negative dynamical corotation torque is really due to the buoyancy resonance, we construct a vertically polytropic disk with $N(z)=0$ (e.g., \citealt{ng2013}), and run simulations with a fixed or moving planet. We take an adiabatic equation of state and do not consider radiative diffusion. 

Figure \ref{fig:buoypoly} plots the vertical velocity perturbations at $z/H=2$ from the fixed-planet model in the polytropic disk at $t=500t_\text{orb}$. Compared to Figure \ref{fig:buoy}, the polytropic disk does not possess structure related to the buoyancy resonance. Figure \ref{fig:plytqwk} plots the temporal variations of the torques on the fixed and moving planets in the polytropic disk. The fact that the torque on the fixed planet is more negative than that on the moving planet indicates that the dynamical corotation torque is positive. This is similar to the behavior in the radiative disk, but is opposite to the case with the adiabatic disk shown in Figure \ref{fig:tqwk}. 

\begin{figure}[t!]
\epsscale{1.2}
\plotone{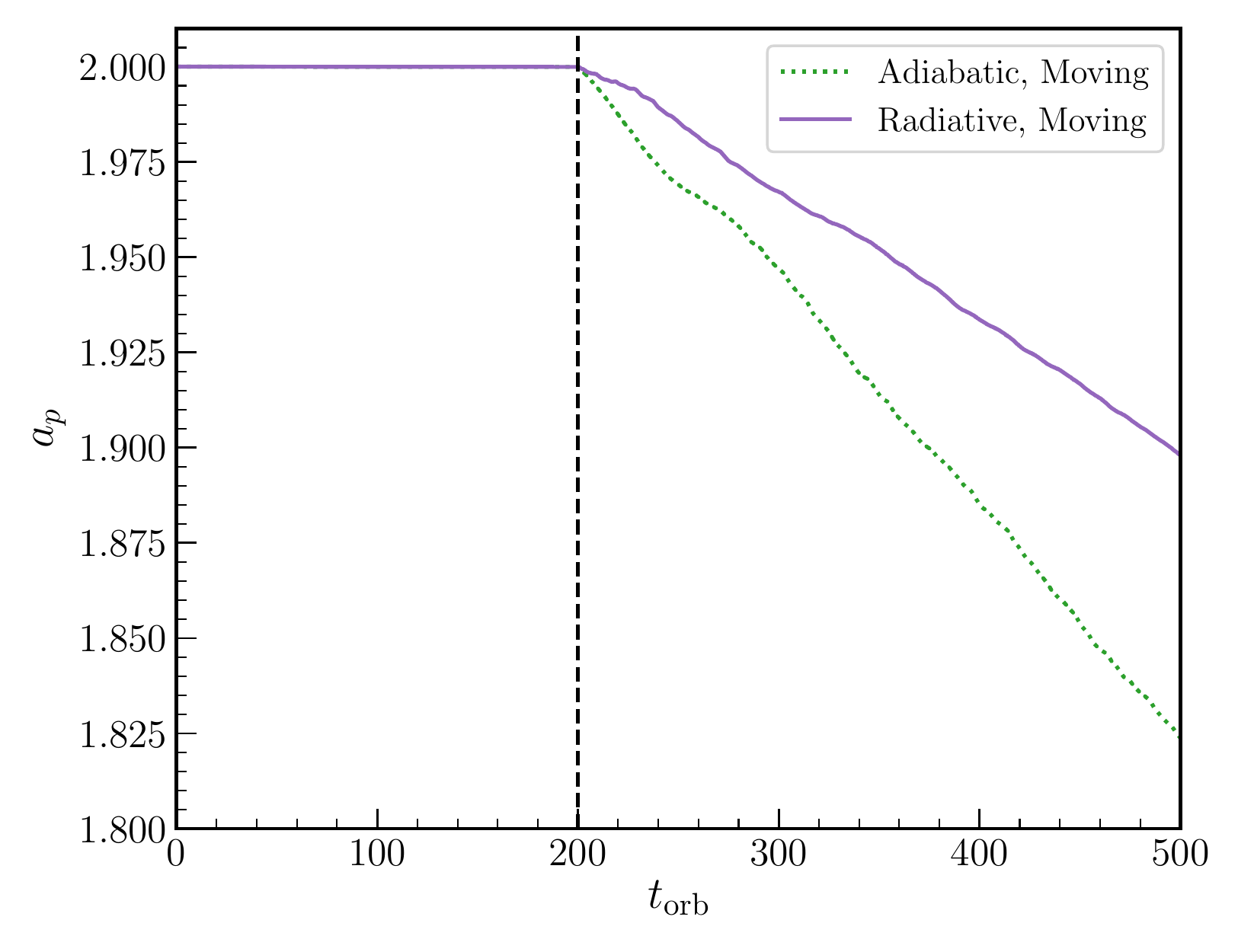}
\caption{Temporal evolution of the semi-major axes $a_p$ of the migrating planet in the adiabatic (\textit{dotted}) and radiative (\textit{solid}) disks. The planet is released at $t = 200t_{\text{orb}}$ (\textit{dashed line}). Note that the radiative diffusion slow downs the inward migration by $\sim57\%$ compared to the adiabatic counterpart. \label{fig:orbit}}
\end{figure}

\citet{mn2020} showed that the buoyancy resonance in a 3D adiabatic disk provides a negative dynamical corotation torque, resulting in a fast inward migration of a planet. However, the radiative diffusion reduces the negative dynamical corotation torque significantly by weakening the buoyancy resonance. Figure \ref{fig:orbit} compares the temporal changes of the semimajor axis $a_p$ of a planet migrating due to the torque it receives in the radiative or adiabatic disk. 
The planet released at $t=200t_\text{orb}$ moves radially in by $\sim0.1a_p$ and $\sim0.175a_p$ over subsequent $300 t_\text{orb}$ in the radiative and adiabatic disks, respectively, showing that the radiative diffusion slows down the inward migration by $57\%$. This suggests that the inclusion of the radiative diffusion helps to alleviate the issue of the rapid migration in adiabatic disks reported by \citet{mn2020}.

Finally, we remark a few caveats of our simulations. First, our model disks have the surface density that decreases monotonically with radius in a power-law fashion $(\Sigma\propto r^{-1/2})$. This precludes a possibility of protoplanet trap that may be created when the surface density has a radial jump \citep{mn2006}. Even with a radial jump in the surface density profile, it would unlikely have a dramatic impact on planet migration in our radiative disk. This is because the protoplanet trap generally results from the dependence of the corotation torque on the local surface density gradient \citep{mn2006, bc2016}, while the corotation torque is saturated in our inviscid radiative disk.
Nonetheless, the surface density profile can still affect the migration of a migrating planet via the dynamical corotation torque which depends on the vortensity gradient in the initial disk.

Second, while our models incorporate radiative diffusion in protoplanetary disks, they still miss a few crucial factors that can make the disks more realistic. The buoyancy resonance is dependent on the Brunt-V\"{a}is\"{a}l\"{a} frequency $N(z)$ which is determined by the vertical disk structure. While we do not consider the irradiation from the central star \citep{bc2013}, it would primarily heat the optically-thin, high-$|z|$ regions to modify $N(z)$, changing the strength of the buoyancy resonance there \citep{bt2021}. Our current models also do not consider the accretion luminosity of an embedded planet. It is well known that by heating the gas close to the planet, the accretion luminosity gives rise to a torque called heating torque \citep{bm2015}. In addition, the density and temperature fluctuations due to the accretion luminosity would alter the behavior of the buoyancy resonance, which can affect the dynamical corotation torque. Moreover, the disk material near the planet leads to a flow instability from the accretion luminosity if the disk has a non-uniform opacity, which can potentially affect the buoyancy resonance as well \citep{cl2019}. Additionally, the disk in our simulation only considers radiative cooling as a cooling process but infrequent collision between gas molecules and dust grains in the surface layers can dominate the thermodynamics in real disk \citep{bt2021}. When infrequent gas-dust collision plays a role, buoyancy resonances will become strong again. To assess the effect of the buoyancy resonance on the planet migration in more realistic situations, therefore, one has to include the stellar irradiation as well as the accretion luminosity and the effect from infrequent gas-dust collision in the thermal evolution of protoplanetary disks.

\section{Conclusion} \label{sec:conc}

The dynamical corotation torque arises from the deformation of the horseshoe orbits along with the vortensity gradient in the background disk. While previous 2D studies on planet migration predicted that the dynamical corotation torque counteracts the inward migration \citep{p2014,mn2019}, the recent 3D simulations of \citet{mn2020} showed that the buoyancy resonance in adiabatic disks can alter the vortensity inside the libration region and change the sign of the dynamical corotation torque for earth-mass planets, causing the planets to migrate even faster. To study how the buoyancy resonance behaves in a more realistic radiative disk, in this paper we include radiative diffusion in an otherwise adiabatic disk, and investigate its effect on the buoyancy resonance and the dynamical corotation torque. We adopt the FLD approximation to realize radiation transfer, and compare the torques from the simulations of the adiabatic and radiative disks with a fixed or moving planet. The main results of the present work can be summarized as follows.

\begin{itemize}
  \item[1.] In 3D adiabatic disks, the dynamical corotation torque is negative, consistent with the results of \citet{mn2020}. 
  
  \item[2.] The radiative diffusion suppresses the buoyancy resonance, making dynamical corotation torque positive. 
  The suppression of the buoyancy resonance is stronger at higher-altitude regions since the diffusion timescale is shorter than the buoyancy timescale there.
  
  \item[3.] In a vertically polytropic disk in which buoyancy resonance is completely absent, the dynamical corotation torque is slightly positive, as well. This confirms that the negative dynamical corotation torque in the 3D adiabatic disk is indeed due to the buoyancy resonance.
  
  \item[4.] The radiative diffusion leads to the formation of cold fingers, consistent with the results of \citet{lc2014}, although their contribution to the total torque is minimal because of the tapering function applied to the torque calculation in our simulations.

\end{itemize}

The radiative diffusion can alleviate the issue of a rapid migration caused by the negative dynamical corotation torque seen in the adiabatic disks of \citet{mn2020}. This highlights the significance of the radiative diffusion on the buoyancy resonance or planet-disk interactions in general. Future studies including stellar irradiation and accretion heating will help fully assess the importance of the buoyancy resonance on the type I migration.

\begin{acknowledgments}
We are grateful to the referee for an insightful report.
This work was supported by the grants of National Research Foundation of Korea (2019R1A2C1004857 and 2020R1A4A2002885) and  Basic Science Research Program through the National Research Foundation of Korea (NRF) funded by the Ministry of Education (2022R1A6A3A13072598). Computational resources for this project were provided by the Supercomputing Center/Korea Institute of Science and Technology Information with supercomputing resources including technical support (KSC-2021-CRE-0088).
\end{acknowledgments}

\appendix
\section{Test of the Radiative Diffusion Module}

We implement the two-temperature FLD method for the radiative diffusion in the FARGO3D as described in Section \ref{sec:method}. To test our implementation, we consider the migration of an accreting protoplanet studied by \citet{cl2019}. 
To make the situation identical to that in \citet{cl2019}, we include the viscous diffusion term with kinematic viscosity $\nu = 10^{15}\unit{cm^2\ s^{-1}}$ as well as the accretion heating term for this test.  The disk is initialized with the surface density $\Sigma = 484 (r/1\unit{au})^{-1/2} \unit{g\ cm^{-2}}$ with constant aspect ratio $h=0.05$ across the disk. 
The disk is radially extended from $r_\text{min}=3.12\unit{au}$ to $r_\text{max}=7.28\unit{au}$ and vertically from $\phi_\text{min}=83^{\circ}$ to $\phi_\text{max}=90^{\circ}$. The disk is resolved by $512\times64\times1382$ evenly spaced cells. We follow \citet{bl1994} to calculate the non-uniform opacity depending on the local gas density and temperature. 

Initially, a planet with mass $M_p=3\unit{M_{\oplus}}$ is forced to move on a circular orbit with the orbital radius of $a_p=5.2\unit{au}$ for the first 30 orbits without any accretion and accretion heating. During this time, the disk adjusts itself to the gravitational potential of the planet which is smoothed  by the tapering cubic-spline function, with the smoothing length taken equal to the half-Hill radius. At $t=30t_\text{orb}$, the planet is allowed to accrete and emit related heat whose luminosity is given by  Equation (7) of \citet{cl2019}, with the 
mass-doubling timescale of $\tau = 100\unit{kyr}$. 

\begin{figure}[h!]
\epsscale{1.0}
\plotone{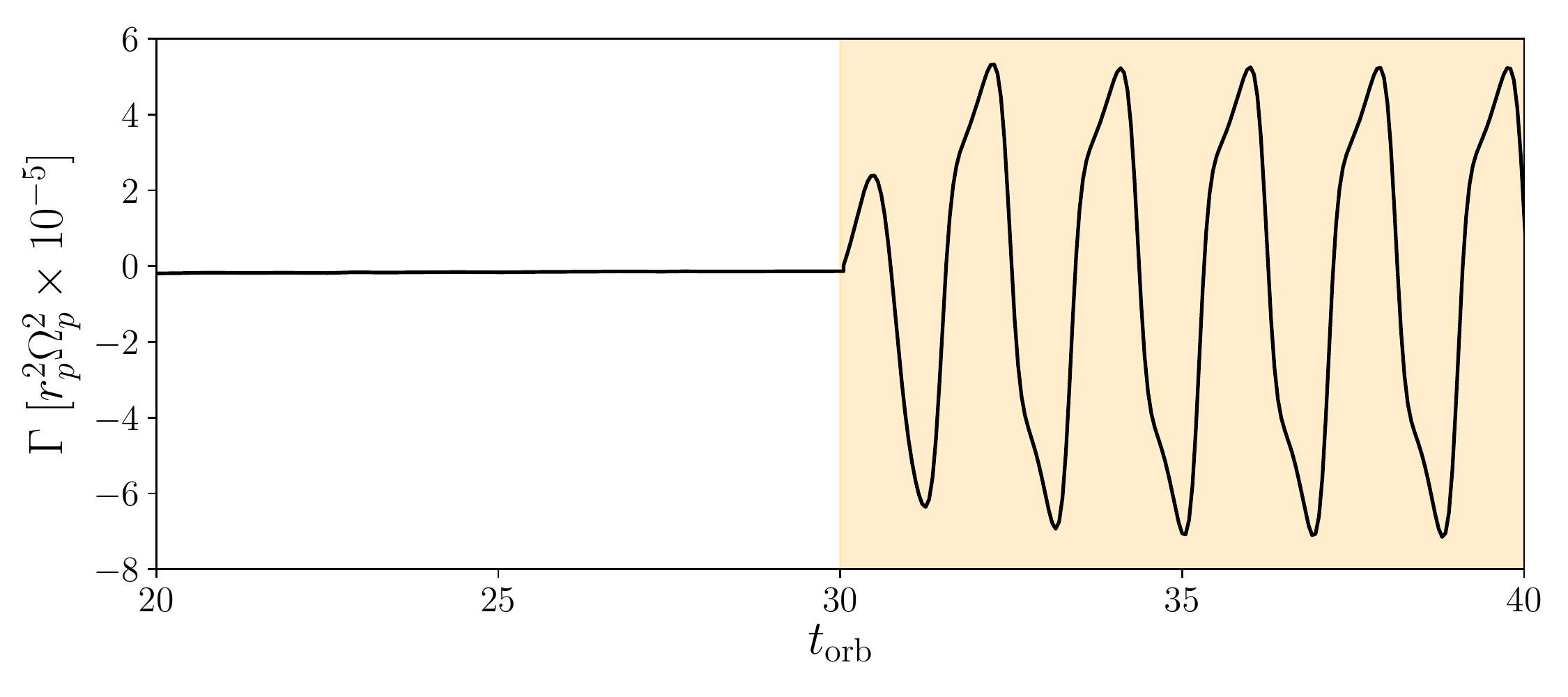}
\caption{Temporal variations of the total torque $\Gamma$ on the accreting protoplanet with mass $M_p=3\unit{M_{\oplus}}$. The accretion and related heating are turned on at $t=30t_\text{orb}$, after which $\Gamma$ exhibits oscillations with a period of $\sim1.9t_{\text{orb}}$ and amplitudes of  $\approx 1.24\times10^{-4}r^2\Omega_p^2$, in good agreement with the results of \citet{cl2019}.\label{fig:osctorq}}
\end{figure}

Figure \ref{fig:osctorq} plots the temporal evolution of the total torque $\Gamma$. The mean torque for $20t_{\text{orb}}<t<30t_\text{orb}$ is $\Gamma=-1.60\times10^{-6}r_p^2\Omega_p^2$, but undergoes strong oscillations which is caused by 3D distortions of the streamlines in the vicinity of the planet. The oscillation has a period  $\sim 1.9 t_{\text{orb}}$ and amplitude $\approx 1.24\times10^{-4}r_p^2\Omega_p^2$, entirely consistent with the results of \citet{cl2019}. 
This proves that our radiative diffusion module reproduces the oscillatory accretion torque in a disk with a non-uniform opacity.

\end{document}